\documentclass[reprint, superscriptaddress, secnumarabic, amsmath, amssymb, nobibnotes, aps, prb]{revtex4-2}

\usepackage{graphicx}
\usepackage{caption}
\usepackage{epstopdf}
\usepackage[T1]{fontenc}
\usepackage{amsbsy}
\usepackage{gensymb}
\usepackage{amsmath}
\usepackage{amssymb}
\usepackage{bbm}
\usepackage{braket}
\usepackage{xcolor}
\usepackage{multirow}
\allowdisplaybreaks
\usepackage{graphicx}
\usepackage[hidelinks]{hyperref}  
\usepackage[normalem]{ulem}
\usepackage{orcidlink}

\renewcommand{\approx}{\simeq}

\begin{document}

	\title{\textrm{A Candidate for the Quantum Spin Liquid Ground-State in the Shastry-Sutherland Lattice Material Yb$_2$Be$_2$GeO$_7$} }
\author{M. Pula\,\orcidlink{0000-0002-4567-5402}}
\email[]{pulam@mcmaster.ca}
\affiliation{Department of Physics and Astronomy, McMaster University, Hamilton, Ontario L8S 4M1, Canada}
\author{S. Sharma\,\orcidlink{0000-0002-4710-9615}}
\affiliation{Department of Physics and Astronomy, McMaster University, Hamilton, Ontario L8S 4M1, Canada}
\author{J. Gautreau}
\affiliation{Department of Physics and Astronomy, McMaster University, Hamilton, Ontario L8S 4M1, Canada}
\author{Sajilesh K. P.}
\affiliation{Physics Department, Technion-Israel Institute of Technology, Haifa 32000, Israel}
\author{A. Kanigel}
\affiliation{Physics Department, Technion-Israel Institute of Technology, Haifa 32000, Israel}
\author{M. D. Frontzek}
\affiliation{Neutron Scattering Division, Oak Ridge National Laboratory, Oak Ridge, Tennessee 37831, USA}
\author{T. N. Dolling\,\orcidlink{0009-0007-9324-1485}}
\affiliation{School of Chemistry, University of Birmingham, Edgbaston, Birmingham B15 2TT, UK}
\author{L. Clark\,\orcidlink{0000-0002-6223-3622}}
\affiliation{School of Chemistry, University of Birmingham, Edgbaston, Birmingham B15 2TT, UK}
\author{S. Dunsiger}
\affiliation{Department of Physics, Simon Fraser University, Burnaby, British Columbia V5A 1S6, Canada}
\affiliation{TRIUMF, Vancouver, British Columbia V6T 2A3, Canada}
\author{A. Ghara}
\affiliation{Department of Physics, Indian Institute of Science Education and Research, Pune, Maharashtra 411008, India}
\author{G.~M.~Luke\,\orcidlink{0000-0003-4762-1173}}
\email[]{luke@mcmaster.ca}
\affiliation{Department of Physics and Astronomy, McMaster University, Hamilton, Ontario L8S 4M1, Canada}
\affiliation{TRIUMF, Vancouver, British Columbia V6T 2A3, Canada}	
	\date{\today}
	\begin{abstract}
		\begin{flushleft}
		\end{flushleft}
The quasi-2D Shastry-Sutherland model has remained topical in the field of condensed matter physics for the last two decades, following the experimental realization of the model in the material SrCu$_2$(BO$_3$)$_2$. Since then, research into the Shastry-Sutherland system has revealed more nuanced physics than initially predicted; recent theoretical works have even predicted a quantum spin liquid phase may exist. Herein, we report on a new Shastry-Sutherland lattice material, Yb$_2$Be$_2$GeO$_7$, of the rare-earth melilite family RE$_2$Be$_2$GeO$_7$. We find, through SQUID magnetometry, powder neutron diffraction, specific heat capacity, and muon spin relaxation, that Yb$_2$Be$_2$GeO$_7$ lacks magnetic order and exhibits persistent spin dynamics to at least 17 mK. We propose the Shastry-Sutherland lattice material Yb$_2$Be$_2$GeO$_7$ as a candidate to host a quantum spin liquid ground-state.
		
	\end{abstract}
	\maketitle
 
\section{Introduction}
A spin liquid (SL) is a state in which the magnetic ordering of interacting spins is suppressed by a macroscopic ground-state degeneracy. This may be introduced via, for example, magnetic frustration, i.e., order is prevented by spins freely fluctuating between unique but degenerate configurations \cite{QSL-tyrel}. Spin liquids occur in two flavors: classical and quantum. In a classical spin liquid, this behavior is thermodynamic in nature, i.e., thermally driven random fluctuation between asynchronous microstates, and occurs only for a finite temperature range \cite{balents2010spin}. Conversely, the quantum spin liquid (QSL) state, as the name suggests, is quantum in nature; fluctuations are driven by the zero-point energy of the uncertainty principle, and spins exist highly entangled and in a superposition of many orientations \cite{balents2010spin}. Such a state is only possible when the zero-point motion is on the order of the size of the spin angular momentum, generally precluding all but S=$\frac{1}{2}$ and S=1 systems \cite{QSL-tyrel,balents2010spin}.

The search for the elusive QSL began in 1973, following the postulations made by Anderson on resonating valence bonds \cite{ANDERSON1973153}. A material that hosts a QSL state has yet to be definitively identified, owing largely to the inability to directly probe the QSL state \cite{balents2010spin}. Nevertheless, various QSL candidate materials (examples in parentheses) have been proposed on several lattice symmetries, including 1D chains (Sr$_2$CuO$_3$ \cite{schlappa2018probing}), honeycomb ($\alpha$-RuCl$_3$ \cite{banerjee2016proximate}), pyrochlore (Yb$_2$Ti$_2$O$_7$ \cite{PhysRevX.1.021002}), diamond (FeSc$_2$S$_4$ \cite{PhysRevLett.92.116401}), triangular (YbMgGaO$_4$ \cite{paddison2017continuous}), kagom\'e (ZrCu$_3$(OH)$_6$Cl$_2$ \cite{PhysRevLett.98.107204}), and hyper-kagom\'e (PbCuTe$_2$O$_6$ \cite{chillal2020evidence}).

QSLs have a number of properties that can be probed (but are by no means definitive proof of a QSL state), such as a lack of magnetic order at all temperatures (which can be confirmed via magnetic susceptibility, specific heat capacity, and elastic neutron scattering measurements) and a persistence of spin fluctuations at T=0 (probeable via inelastic neutron scattering, muon spin relaxation/rotation ($\mu$SR) and nuclear magnetic resonance). QSLs may also fractionalize electrons into holons and spinons, and the associated spinon excitation can be seen in inelastic neutron scattering (which forms a continuum \cite{han2012fractionalized}) and spinon thermal conductivity (if the spinons are itinerant).

\begin{figure}
    \centering
    \includegraphics[width=\columnwidth]{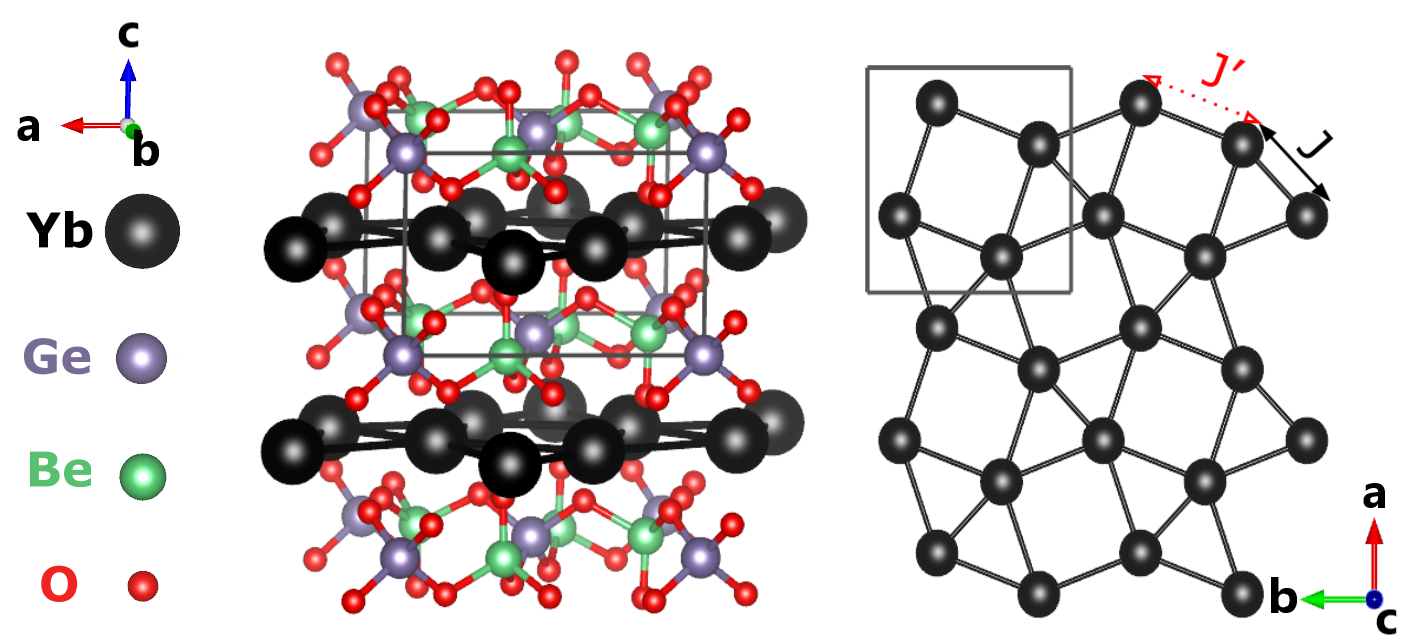}
    \caption{\textbf{Left}: The crystal lattice of Yb$_2$Be$_2$GeO$_7$, which has the space group P$\overline{\mbox{4}}$2$_1$m. The unit cell is shown in dark grey. \textbf{Right}: AB-Planar arrangement of the RE$^{3+}$ atoms. Here, $J$ and $J^{'}$ denote the nearest-neighbor and next-nearest-neighbor interactions. If both $J$ and $J^{'}$ are antiferromagnetic, the lattice is expected to be geometrically frustrated. The unit cell is shown in dark grey.}
    \label{fig:Re2Be2GeO7-lattice}
\end{figure}

Recently, M. Ashtar, Y. Bai \emph{et~al.}~\cite{doi:10.1021/acs.inorgchem.0c03131} synthesized members of a family of rare-earth (RE) materials isomorphic with melilite, RE$_2$Be$_2$GeO$_7$. These rare-earth melilites form in alternating layers of non-magnetic Be$_2$GeO$_7$ and magnetic RE$^{3+}$ ions and lack detectable site mixing of the constituent ions. The RE$^{3+}$ ions form with an in-plane geometry topologically equivalent to the Shastry-Sutherland lattice (SSL) (see Fig.~\ref{fig:Re2Be2GeO7-lattice}), a theoretical model proposed by Shastry and Sutherland \cite{SRIRAMSHASTRY19811069} as a 2D Heisenberg Hamiltonian with antiferromagnetic nearest ($J$) and next-nearest neighbor ($J^{'}$) interactions. The SSL is predicted to host a spin-dimer state when $J$>>$J^{'}$ and a square antiferromagnet when $J^{'}$>>$J$; the specific transition between these two phases was calculated by Miyahara and Ueda to be $J^{'}$/$J$ $\approx$ 0.7 \cite{PhysRevLett.82.3701}. The SSL is well-known as the first 2D spin system to exhibit fractional magnetization plateaus \cite{PhysRevLett.82.3168}, observed in the first known SSL material SrCu$_2$(BO$_3$)$_2$.

\begin{figure*}
      \centering
    \includegraphics[width=\linewidth]{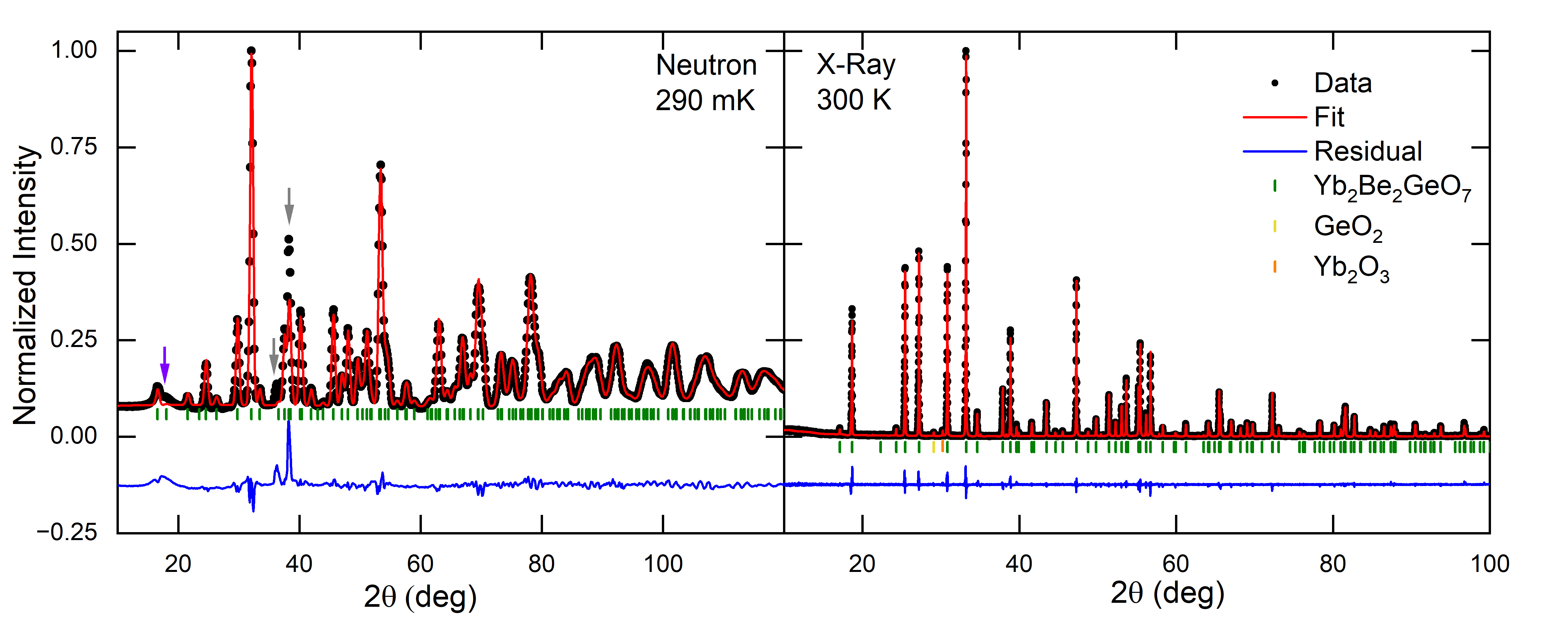}
    \caption{\textbf{Left}: Neutron (Ge (113) $\lambda$=1.486~\AA) \textbf{Right}: X-Ray (Cu K$_{\alpha1}$ $\lambda=1.541$~\AA)  elastic scattering of Yb$_2$Be$_2$GeO$_7$. The green marks indicate the main phase, Yb$_2$Be$_2$GeO$_7$. The orange and red marks indicate impurity peaks, only detectable in the XRD pattern, attributed to  GeO$_2$ and Yb$_2$O$_3$. In the PND pattern, the Purple arrow around 18$\degree$ indicates scattering due to FC-40. Grey arrows correspond to the (002) and (102) peaks, which over-contribute. The Rietveld refinement was performed on FullProf; see Table \ref{Tab:Rietveldpara} of Appendix B for a summary the fitting parameters.}
    \label{fig:elastic}
\end{figure*}

Contemporary experiments \cite{guo2020quantum,jimenez2021quantum} on SrCu$_2$(BO$_3$)$_2$ have revealed evidence of a plaquette state, accessible via pressure-induced tuning of $J^{'}$/$J$, residing in the intermediate region between the spin-dimer and antiferromagnetic phases. This has inspired a number of theoretical works into the SSL system, which predicts that a spin liquid phase is possible for a narrow window of $J^{'}$/$J$ ratios \cite{Wang_2022, PhysRevB.105.L041115, PhysRevB.105.L060409}.

In this paper, we report our findings on a particular member of the rare-earth melilites: the Yb-based rare-earth melilite Yb$_2$Be$_2$GeO$_7$. We find that, despite the presence of magnetic interactions, Yb$_2$Be$_2$GeO$_7$ lacks any signature of magnetic ordering and has persistent spin dynamics down to 17 mK. We, therefore, propose the Shastry-Sutherland system Yb$_2$Be$_2$GeO$_7$ as a candidate for the QSL ground-state under ambient pressure.

\section{Methods}

Polycrystalline Yb$_2$Be$_2$GeO$_7$  was synthesized via solid-state reaction, following the methodology of Y. Ochi \emph {et~al.} \cite{OCHI1982911}, as well as Y. Bai and M. Ashtar \cite{doi:10.1021/acs.inorgchem.0c03131}. Stoichiometric ratios of Yb$_2$O$_3$ (4N), BeO (4N), and GeO$_2$ (4N) powders were mechanically ground/mixed under an argon environment, then fired at 1350$\degree$C for 24h; this process was repeated four times. The resulting powders were pelleted, then sintered at 1350$\degree$C for 24h. 

\begin{figure}
      \centering
    \includegraphics[width=\columnwidth]{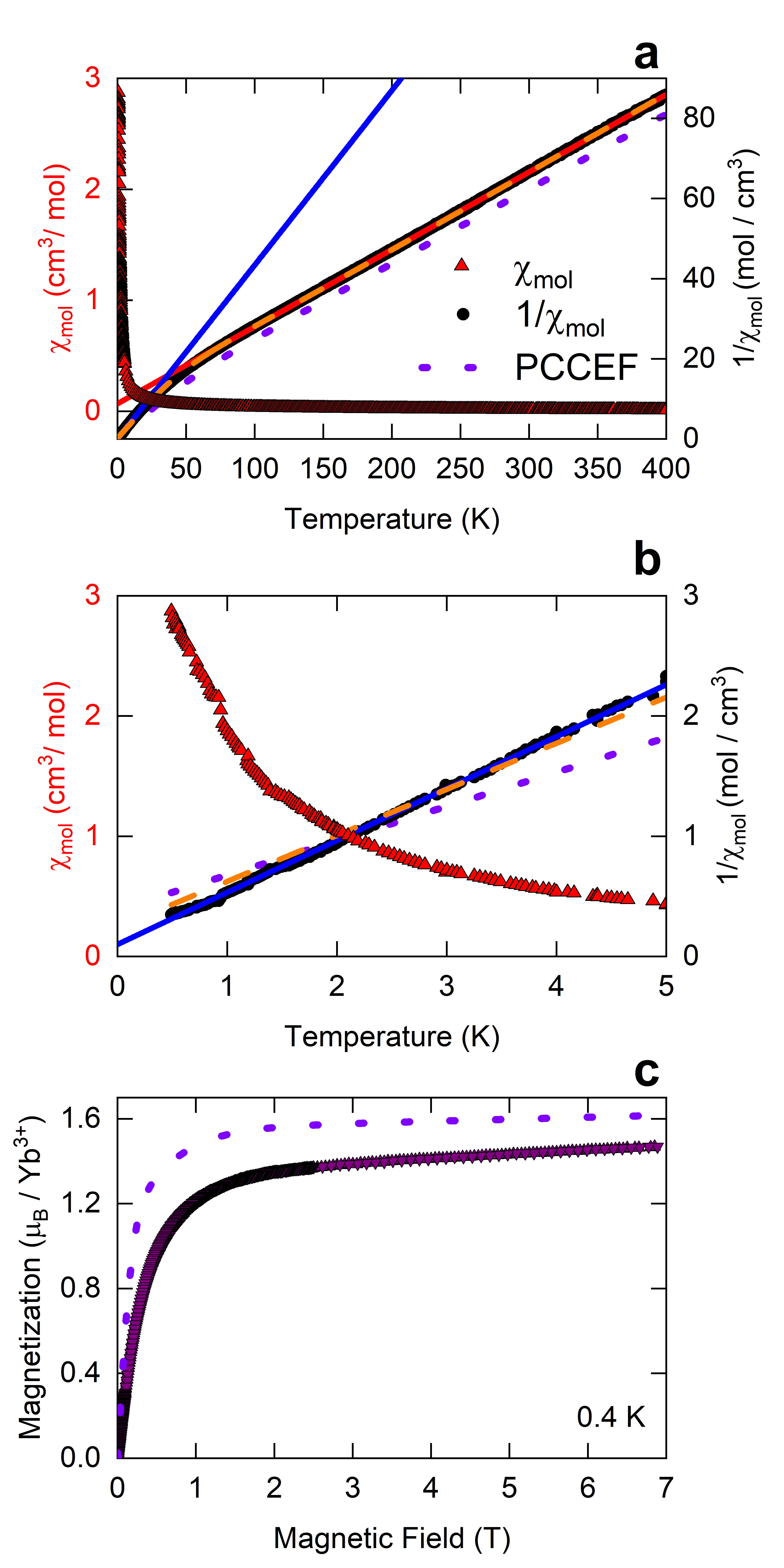}
    \caption{Magnetometry measurements of Yb$_2$Be$_2$GeO$_7$. \textbf{a}: $\chi$(T) with two-level Curie-Weiss fit (orange dashed line). The red and blue solid lines represent conventional Curie-Weiss fits of the high-and-low-temperature regimes, respectively. The purple dotted line corresponds to susceptibility as predicted by the PCCEF calculations. \textbf{b}: $\chi$(T) at low temperatures with two-level (orange dashed line) and conventional (blue solid line) Curie-Weiss fits. PCCEF susceptibility is shown as the dotted purple line. \textbf{c}: Magnetization at 0.4 K. The dotted purple line represents the magnetization as predicted by the PCCEF calculations.}
    \label{fig:SQUID}
\end{figure}

The sample composition was confirmed by X-ray diffraction (XRD), which was performed on a Panalytical X'pert Pro diffractometer. Isothermal magnetization (i.e., Fig.~\ref{fig:SQUID}c) measurements were collected with a Quantum Design MPMS3 magnetometer with a $^3$He insert on a sample of mass~9.2~mg. Magnetic susceptibility measurements were performed using a Quantum Design MPMS XL magnetometer. We conducted measurements in the temperature range from 1.8~K to 400~K with a 0.2~T field on a sample of mass 58.0~mg. Low-temperature measurements were performed with an iQuantum $^3$He insert in the temperature range from 0.49~K to 4.8~K with a 0.02~T field on a 5.31~mg sample. These data sets are combined in Figs.~\ref{fig:SQUID}a and 3b. We carried out powder neutron diffraction (PND) measurements using a vanadium sample can loaded with loose Yb$_2$Be$_2$GeO$_7$ powder and Fluorinert (FC-40) on the Wide Angle Neutron Diffractometer (WAND$^2$) at Oak Ridge National Laboratory's High Flux Isotope Reactor. Specific heat capacity measurements employed the two-tau relaxation method on a Quantum Design PPMS equipped with a dilution refrigerator. La$_2$Be$_2$GeO$_7$ was used as a non-magnetic analog to remove the lattice contribution to the specific heat. Our $\mu$SR measurements were performed at TRIUMF (Canada's Particle Accelerator Centre) using the M20 and M15 muon beamlines of the Centre for Molecular and Material Science. Measurements performed on M15 were performed on the Pandora spectrometer with a dilution refrigerator cryostat; a single sintered sample pellet was mounted on a silver coldfinger, thermally coupled via a 50/50 mixture of Cryocon and Apiezon N greases, and wrapped in a thin 99.95$\%$ Ag foil. The M20 measurements used the LAMPF spectrometer with a helium gas flow cryostat. Sintered sample pellets were mounted via Mylar tape on a low background probe; muons missing the sample were detected in a veto counter and not recorded in the measured spectra. 

\begin{table}
\begin{ruledtabular}
    \begin{tabular}{c|cccc}
    Method & Range (K) & $\mu_{eff}$ ($\mu_B$)& $\theta$ (K) & $\chi_0$ (cm$^3$/mol) \tabularnewline
    \hline\
    High-T& 125 $\rightarrow$ 400 & 4.56(3)& -46.2(9) & -7(5)$\times$10$^{-5}$
    \tabularnewline
    Low-T& 0.49 $\rightarrow$ 15 & 3.034(5) & -0.236(7) & fixed
    \tabularnewline
    Two-level& 0.49 $\rightarrow$ 400 &3.232(6), & -0.63(4) & fixed \tabularnewline
    & & 5.525(3) & &
    \end{tabular}
\end{ruledtabular}
\caption{Summary of the Curie-Weiss parameters for the conventional Curie-Weiss fitting of the high-and-low temperature regimes, as well as the two-level model. Both effective moments $\mu_1$ and $\mu_2$ are listed, in that order, for the two-level model.}
\label{tab:CurieWeissPara}
\end{table}

\section{results}

\begin{figure}
      \begin{center}
    \includegraphics[width=\columnwidth]{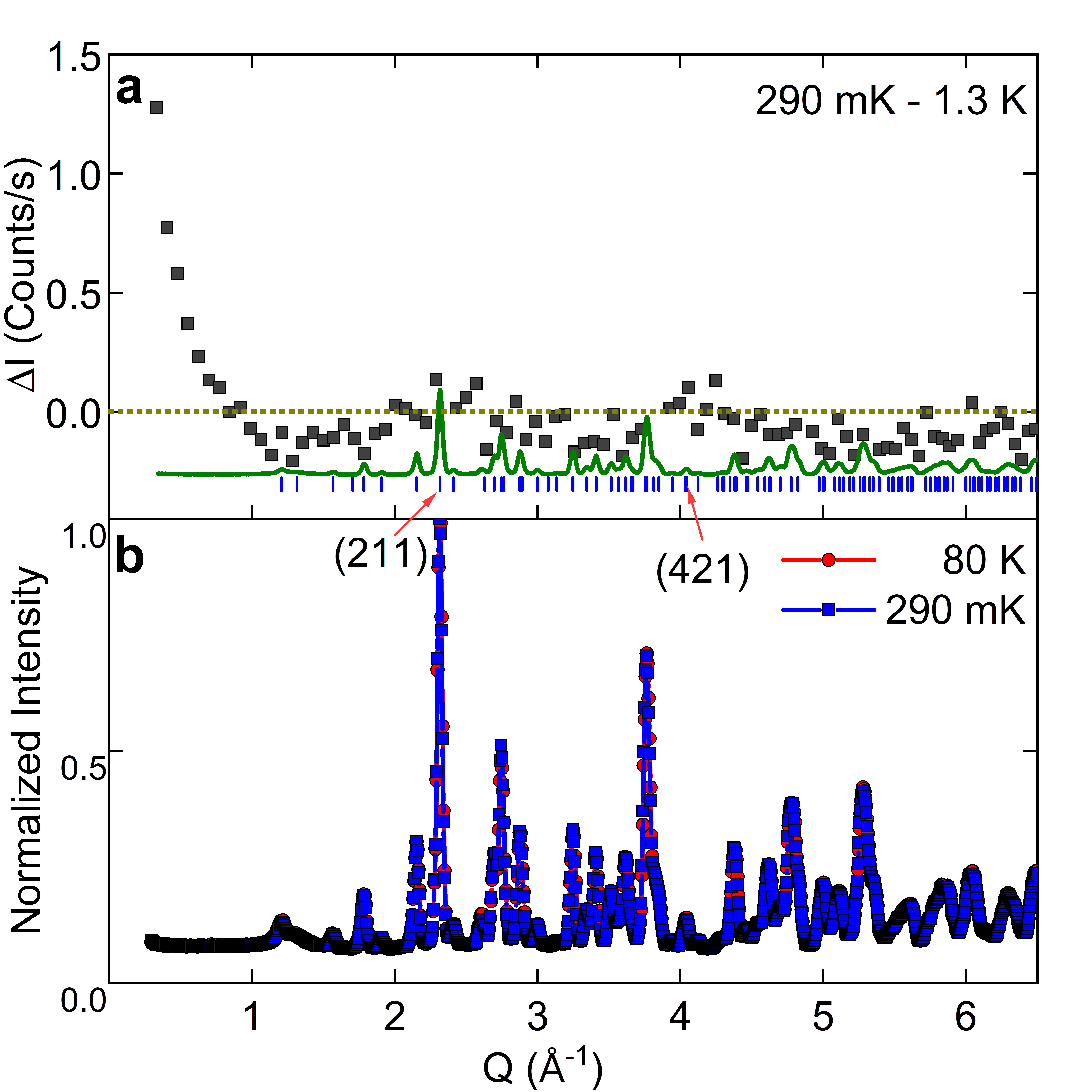}
    \caption{Elastic neutron scattering of Yb$_2$Be$_2$GeO$_7$. \textbf{a}: The difference between patterns collected at 290~mK and 1.3~K. The green line is the PND pattern collected at 290~mK, indicating the locations of the experimentally allowed crystallographic Bragg peaks. The blue ticks correspond to allowed crystallographic Bragg peaks, determined via Rietveld refinement of the aforementioned 290~mk PND pattern. Broad peaks are observed around Q~=~0, 2.3, and 4.1~\AA$^{-1}$. The Q~=~0 peak may indicate ferromagnetic SRO. \textbf{b}: Normalized patterns at 80~K and 290~mK. No qualitative difference in patterns is discernible, indicating the absence of magnetic Bragg peaks down to 290~mK.}
    \label{fig:ENS}
    \end{center}
\end{figure}

The X-ray and powder neutron diffraction patterns of Yb$_2$Be$_2$GeO$_7$ are shown in Fig.~\ref{fig:elastic}. The patterns correspond well with the expected tetragonal space group P$\overline{\mbox{4}}$2$_1$m (113). Rietveld refinement results in the lattice parameters of a~=~b~=~7.33557(2)~\AA \space and c~=~4.75520(2)~\AA \space for X-rays at 300~K, and a~=~b~=~7.3005(9)~\AA \space and c=4.7351(6)~\AA \space for neutrons at 290~mK. A summary of the refinement parameters can be found in Table \ref{Tab:Rietveldpara} of Appendix B. There are slight impurity phases present in the X-ray diffraction pattern, with peaks centered at 29.1$\degree$ and 30.2$\degree$; these are attributed to GeO$_2$ and Yb$_2$O$_3$, which have maximum intensity peaks around 29$\degree$ \cite{baur1956verfeinerung} and 30$\degree$ \cite{saiki1985structural}, respectively. These phases are included in the Rietveld refinement of the XRD pattern. The additional peaks that would be associated with GeO$_2$ and Yb$_2$O$_3$ are undetected, presumably due to the insignificant molar fraction of these impurity phases. GeO$_2$ is diamagnetic \cite{LandoltBornstein2007} and therefore inconsequential to the magnetic properties. Yb$_2$O$_3$ is known to be antiferromagnetic, with a N\'eel temperature of 2.3K \cite{moon1968magnetic}. However, no evidence of this antiferromagnetic transition is observed in SQUID magnetometry, PND, nor specific heat capacity, indicating a negligible contribution to the bulk properties of the sample. In the PND pattern, the (002) and (102) peaks over-contribute to the pattern; this is not well captured by the refinement and can not be attributed to GeO$_2$ nor Yb$_2$O$_3$, which have only weakly scattering peaks in the vicinity. This feature is present up to the highest measured temperature of 80~K (see Fig. \ref{fig:ENS}b), hence it is unlikely to be magnetic in nature.

Magnetic susceptibility in temperature $\chi$(T)  measurements reveal no sign of magnetic ordering down to 500~mK (see Fig.~\ref{fig:SQUID}b). Notably, $\chi$ (T) was observed to have two distinct slopes in 1/$\chi$ that intersect around 40~K (see Fig.~\ref{fig:SQUID}a), an observation shared by Ashtar \& Bai. This behavior is likely due to crystal electric fields and is well modeled by a two-level Curie-Weiss law \cite{mugiraneza2022tutorial, Mitric_1997, BESARA201423}

\begin{equation}
    \frac{1}{\chi - \chi_0} = \frac{3 k_B}{N_A}  (T - \theta) \frac{(1 + e^{-\frac{E_{1}}{k_B T}})}{\mu_{1}^2 + \mu_{2}^2 e^{-\frac{E_{1}}{k_B T}})}
    \label{eqn:CWLawTwoLevel}
\end{equation}
where $\chi_0$ is the temperature independent susceptibility, N$_A$ is Avogadro's number, $k_B$ is Boltzmann's constant, $\theta$ is the Weiss parameter, E$_1$ is the energy separation between the ground-state and the first-excited-state, and $\mu_{1,2}$ are the effective moment numbers of the ground-state and first-excited-state, respectively. This relation is, in essence, the nominal Curie-Weiss law, with the temperature-independent Curie constant, C, replaced by a temperature-\textit{dependent} function, determined via Fermi-Dirac statistics of a two-level system. Note that to avoid over-fitting, the temperature-independent term in eq.~\ref{eqn:CWLawTwoLevel} was determined to be $-7(5)\times10^{-5}$~cm$^3$/mol by a high-temperature one-level (the conventional Curie-Weiss law) fit from 125 K to 400 K and fixed to this value.

The use of the two-level model results in parameters of $\mu_1$~=~3.232(6)$~\mu_B$, $\mu_2$~=~5.525(3)$~\mu_B$, E$_1$~=~142.8(8)~K, and $\theta$~=~-0.63(4) K. In the high-temperature limit (k$_B$T>>E$_1$), Eq.~\ref{eqn:CWLawTwoLevel} predicts an effective moment of 4.53$~\mu_B$. Yb$^{3+}$ has a one-less-than-full 4f valence shell; thus, it is expected by Hund's rules to occupy a state with S= $\frac{1}{2}$ and L = 3, giving a free ion moment of 4.54$~\mu_B$. In the low-temperature limit, the effective moment is simply $\mu_0$ =3.232(6)$~\mu_B$. Conventional Curie-Weiss fits in the high- and low-temperature regimes yield 4.56(3)$~\mu_B$ and 3.034(5)$~\mu_B$. The magnetization of Yb$_2$Be$_2$GeO$_7$ reaches a value of 1.47$~\mu_B$/Yb$^{3+}$ in a field of 7~T at 0.4 K. A summary of the Curie-Weiss parameters can be found in Table \ref{tab:CurieWeissPara}.

For insight into the single-ion effects, we utilized the point charge crystal electric field (PCCEF) model. Although the true CEF levels are only determinable via experiment, the bulk magnetic properties predicted by the PCCEF model both qualitatively agree with our experimental results and have reasonable quantitative agreement (see Figs. \ref{fig:SQUID}.a and \ref{fig:SQUID}.c). Details of this model can be found in Appendix A.

Unlike the case of the dimer singlet SrCu$_2$(BO$3$)$_2$, the susceptibility does not show any indication of approaching zero at very low temperatures, nor is a maximum observed (which should occur at $\sim$0.6 K for the gap of 0.840(3)~K deduce via the specific heat capacity). Interestingly, conventional Curie-Weiss analysis produces a Weiss parameter of almost [-0.236(7) K] zero. A Weiss parameter of zero implies that such a system is purely paramagnetic. However, the isomorphic Tb$_2$Be$_2$GeO$_7$ is known to order antiferromagnetically at T$_N$=2.5 K \cite{doi:10.1021/acs.inorgchem.0c03131}, hence magnetic interactions are expected. Additionally, diffuse magnetic scattering is observed, which will be discussed later. As such, it is unlikely that Yb$_2$Be$_2$GeO$_7$ is purely paramagnetic; the trivial Weiss parameter may be artificially due to competing ferromagnetic and antiferromagnetic interactions.

Next, we present the results of powder neutron diffraction. Figure~\ref{fig:ENS}.b shows the normalized spectra at 80~K and 290~mK; they are quantitatively identical. We know from SQUID magnetometry that Yb$_2$Be$_2$GeO$_7$ is paramagnetic down to $\sim$500~mK, hence the PND pattern at 80~K shows elastic scattering from the nuclear sites of the crystal lattice only.
Since no perceivable additional Bragg peaks emerge at 290 mK, we conclude that Yb$_2$Be$_2$GeO$_7$ does not magnetically order down to 290~mK, which corroborates the lack of long-range magnetic order seen in magnetometry. Figure~\ref{fig:ENS}.a shows the difference between the patterns at 290 mK and 1.3 K. Again, no magnetic Bragg peaks were detected. However, diffuse scattering is observed as broad peaks centered at Q~=~0, 2.3, and 4.1~\AA$^{-1}$. Away from the broad peaks, the intensity observed in the 290 mK pattern is less than that of the 1.3 K pattern, hence counts are consumed by the diffuse scattering in the 290 mK pattern. Diffuse peak maximums at 2.3 and 4.1~\AA$^{-1}$ are closest to the (211) and (421) crystallographic diffraction peaks, respectively. However, the unresolvable lineshape interferes with determining an accurate location in Q-space. The diffuse scattering near Q = 0 implies that short-range ferromagnetic correlations are present. This is somewhat puzzling since this precludes geometric frustration on the SSL and hence magnetic order is expected. In contrast, bulk magnetometry measurements imply that antiferromagnetic interactions are present via the negative Curie-Weiss temperature. The observed diffuse scattering may be a combination of ferromagnetic and antiferromagnetic correlations. 
\newline

\begin{figure*}
    \centering
    \includegraphics[width=\textwidth]{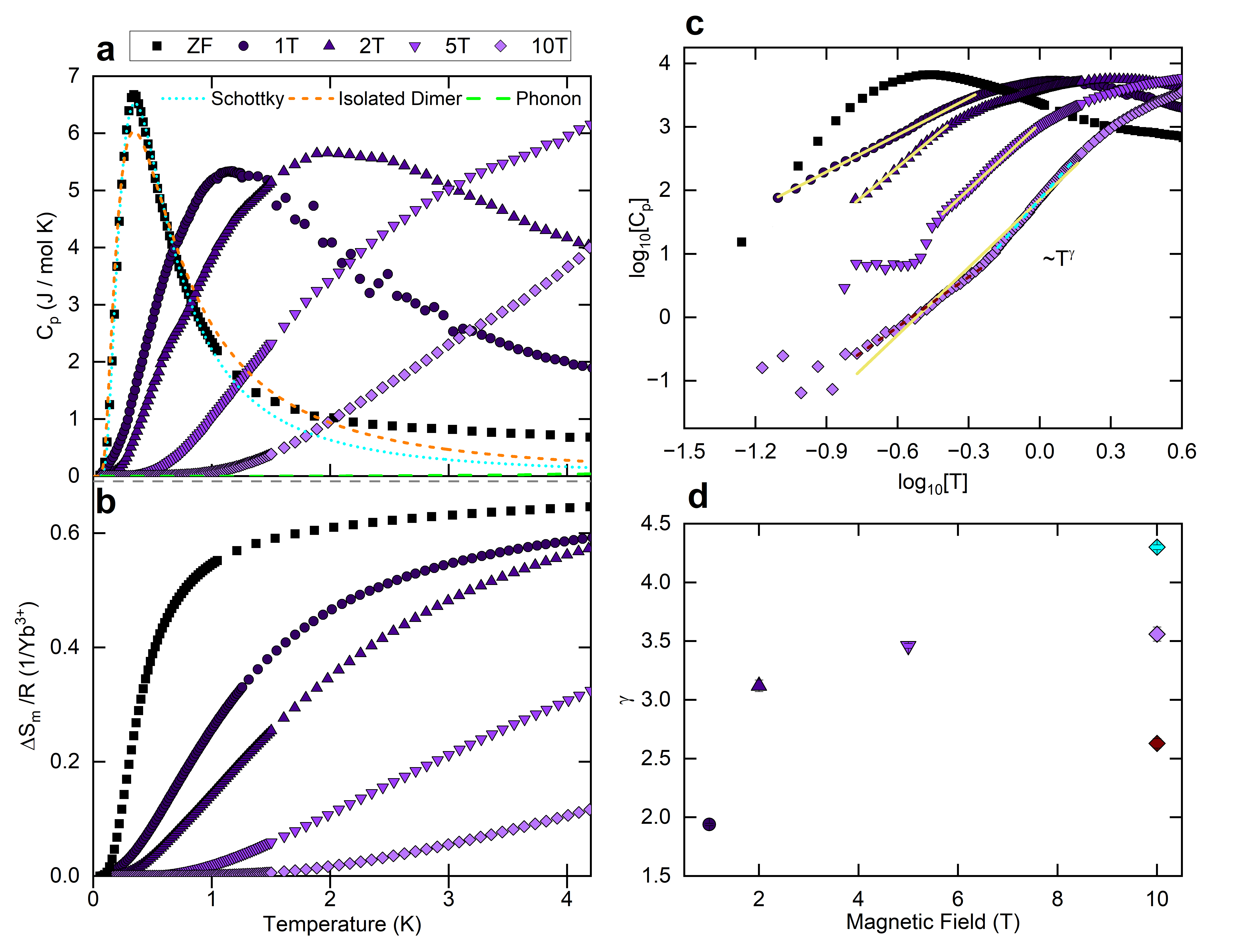}
    \caption{\textbf{a}: Specific heat capacity of Yb$_2$Be$_2$GeO$_7$. Both the two-level Schottky and isolated dimer models are fit from 55 mK to 1 K. The phonon contribution, estimated by the non-magnetic analog La$_2$Be$_2$GeO$_7$, is denoted by the green dashed line. \textbf{b}: Entropy observed in Yb$_2$Be$_2$GeO$_7$, calculated as $S=\int \frac{C_p}{T} dT$. The grey dashed line corresponds to ln(2). \textbf{c}: Log-log plot of the specific heat capacity. The yellow lines correspond to the power-law fits at each field. The cyan and red dashed lines represent the individual power-law fits for the two distinct slopes seen above and below 0.5~K in the 10~T field. \textbf{d}: Power-law exponent $\gamma$ at various the fields. The cyan and red points at 10~T represent the slopes above and below 0.5 K, respectively.}
    \label{fig:specific heat}
\end{figure*}

Our heat capacity measurements, again, confirm a lack of any phase transition (see Fig.~\ref{fig:specific heat}a) down to 55~mK. In zero field, a broad peak is observed, with a maximum of around 0.35~K, which resembles the characteristic shape of a Schottky anomaly. Two-level Schottky anomalies have the mathematical form \cite{carlin2012magnetochemistry}:

\begin{equation}
    C = nAR( \frac{\Delta}{T})^2 \frac{e^{-\frac{\Delta}{T}}}{(1+Ae^{-\frac{\Delta}{T}})^2}
    \label{eqn:Schottky}
\end{equation}
Where C is the specific heat capacity, R is the gas constant, n is the occupancy, $\Delta$ is the energy difference between the two levels, and A is an integer representing the degeneracy. This relation follows simply by considering the partition function of a two-level system is:

\begin{equation}
    Z = 1 + A e^{-\frac{\Delta}{T}}
\end{equation}
The specific heat capacity is shown in Fig.~\ref{fig:specific heat}a, along with a fit to a two-level Schottky model (Eq.~\ref{eqn:Schottky} with A=1, the cyan dotted line in Fig.~\ref{fig:specific heat}a). This model results in an energy gap of 0.840(3) K and an occupancy of n=0.91(1), indicating the majority of Yb$^{3+}$ ions are accounted for. This model fails to capture the specific heat capacity above $\sim$1.5~K, which decays more slowly with increasing temperature than predicted.

The specific heat capacity of the SSL material SrCu$_2$(BO$_3$)$_2$ \cite{kageyama2000specific, kageyama2000low} is known to be well modeled by an isolated dimer model  \cite{carlin2012magnetochemistry, kageyama2000low}. For Yb$_2$Be$_2$GeO$_7$, this is probably not the case. Along with the limitation above 1.5 K seen with the two-level Schottky model, the isolated dimer model (Eq.~\ref{eqn:Schottky} with A=3 for the three triplet states, orange dashed line in Fig.~\ref{fig:specific heat}.a) does not adequately capture the peak profile around its maximum. However, a two-level nondegenerate system could still describe a dimer model, albeit one in which the degeneracy of the triplet levels is lifted.

With the application of a magnetic field, the maximum of the heat capacity shifts to higher temperatures, reaching 1.1 K in a 1 T field, and the profile of the peak is significantly broadened. Furthermore, the behavior approaching T $\rightarrow$ 0 is no longer exponential, likely indicating the closing of the gap in this field. Interestingly, the low-temperature specific heat capacity takes on a power-law dependence in temperature, evident in a log-log plot (Fig.~\ref{fig:specific heat}c), at this field and remains as a power-law, with a field-dependent exponent, for larger fields. This behavior indicates that the gap remains closed with an increasing field.  In a field of 10~T, two distinct slopes [2.58(4) and 4.27(1)] are observed in the logarithmic plot of the specific heat capacity versus temperature, with the slope change occurring around T~$\approx$ 0.5 K. The exponent of the power law for each measured field is shown in Fig.~\ref{fig:specific heat}d.

\begin{figure*}
    \centering
    \includegraphics[width=\textwidth]{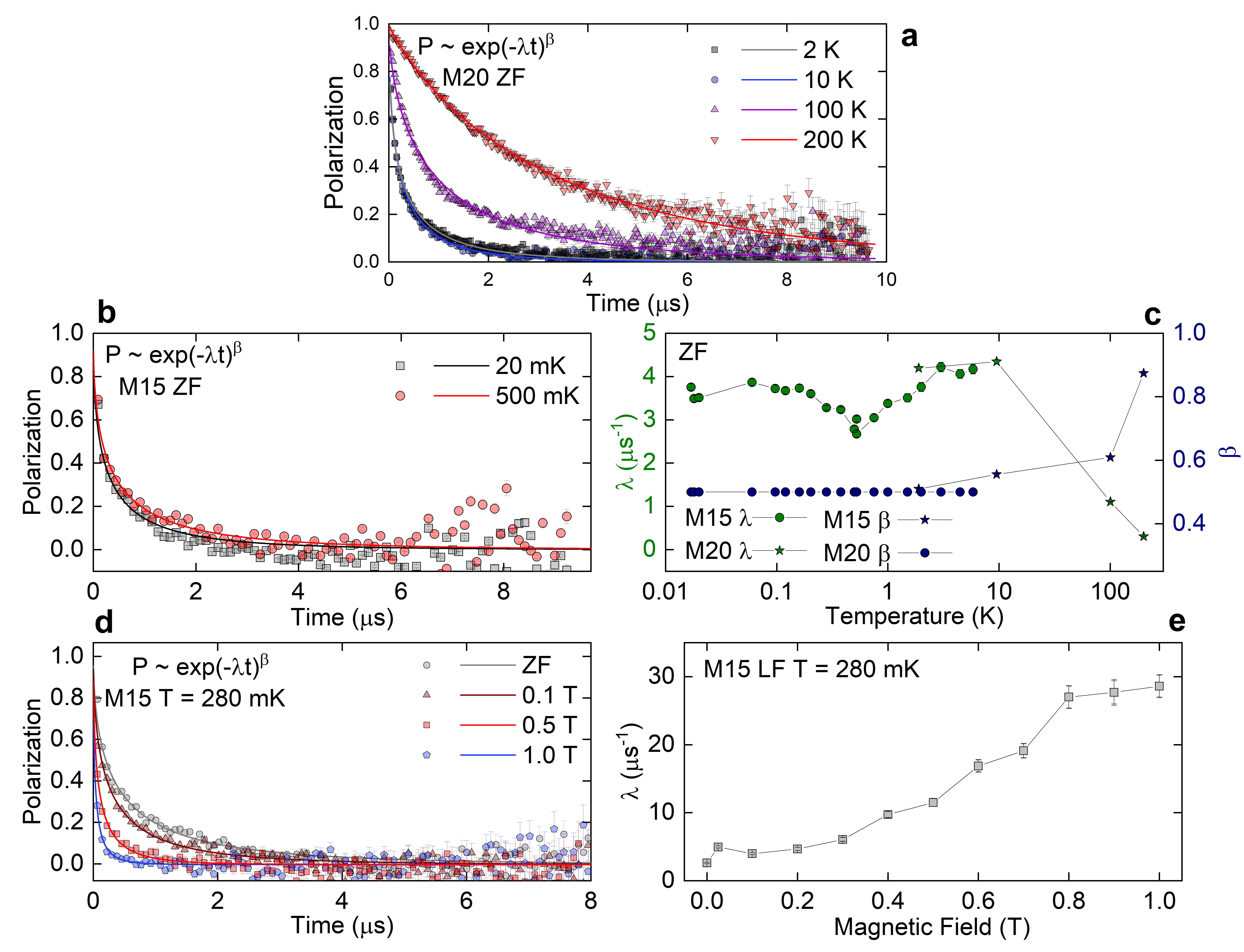}
    \caption{$\mu$SR on Yb$_2$Be$_2$GeO$_7$ powder. \textbf{a}: Raw zero-field spectrum collected with a low-background probe on M20. The data is fit (solid lines) according to Eq.~\ref{Eqn:M20Depol}. \textbf{b}: Raw zero-field spectrum collected with a silver cold-finger on M15. The data is modeled (solid lines) via Eq.~\ref{Eqn:M15Depol}.  \textbf{c}: Fitting results of the relaxation rate $\lambda$ and exponential stretching exponent $\beta$ with temperature. \textbf{d}: Longitudinal-field spectra collected for Y$_2$Be$_2$GeO$_7$ mounted on a silver cold-finger (on the beamline M15). Fits to the spectra determined by Eq.~\ref{Eqn:M15Depol} are shown as solid lines. \textbf{e}: Fitting results for $\lambda$ in the applied longitudinal fields.}
    \label{fig:muSR}
\end{figure*}

The entropy in zero field (Fig.~\ref{fig:specific heat}b) almost reaches Rln2 by 4.2 K. This is consistent with an effective spin $\frac{1}{2}$ system, a result that is also predicted by our PCCEF calculations (see Appendix A). No significant residual entropy is observed, hence magnetic order is not expected to occur below the lowest measured temperature. With the application of a magnetic field, the low-temperature entropy is significantly decreased (down to 0.12 R at 10 T). This is likely simply due to the location of the maximum in the specific heat capacity shifting to higher temperatures. Since the integrand of the entropy scales as 1/T and the temperature range is set by experimental limits, much of the change in entropy occurs outside the experimental temperature window for fields~>~5~T. \newline

Lastly, we will discuss our muon spin rotation/relaxation ($\mu$SR) data. In $\mu$SR, a homogeneous internal field results in a sinusoidal oscillation in the muon polarization function (see, e.g. \cite{motla2021type}) in the time spectrum, which can indicate conventional magnetic order \cite{kojima1997reduction}. No homogeneous fields were detected in Yb$_2$Be$_2$GeO$_7$, however, depolarization of the muons was present.

Depolarization can occur due to stochastic static field distributions (e.g. nuclear dipole moments \cite{sharma2021fully, sharma2023evidence}, spin freezing \cite{sharma2022synthesis}), incommensurate magnetism, inhomogeneous spin densities, or dynamic field distributions (from spin fluctuations). Longitudinal field $\mu$SR can be used to differentiate stochastic static and dynamic field distributions, hence the fluctuation rate of spins in dynamical depolarization processes is typically quite large, while the gyromagnetic ratio of muons is comparatively small. As a result, dynamical depolarization processes tend to be field-independent until relatively large fields are applied. This is not the case for static random fields, whose associated depolarization can be prevented, a process known colloquially as decoupling, by applying a field on the order of the width of the field distribution describing them \cite{uemura1999musr}. Additionally, the polarization function is depolarized to zero for dynamical fluctuations. This is not the case for polycrystalline specimens with stochastic static fields, where $\frac{1}{3}$ of the polarization will be preserved in the absence of an applied (longitudinal) field. This preserved $\frac{1}{3}$ polarization component corresponds to the fraction of muons that experiences a local field parallel to their initial spin direction and hence are not depolarized. 

The polarization functions in zero field $\mu$SR, performed on TRIUMF's M15 beamline (see Fig. \ref{fig:muSR}.b) in the temperature range of 17 mK to 6 K, were modeled phenomenologically with a stretched exponential function:

\begin{equation}
    P = (F e^{-( \lambda t) ^{\beta}} + (1-F)e^{- \lambda_{AG} t })
    \label{Eqn:M15Depol}
\end{equation}
Where $\lambda$ is the relaxation rate, $\beta$ is the stretching exponent, and F is the sample fraction. The second term in the expression represents the contribution of the silver background, which was found to have a small relaxation of  $\lambda_{AG}$ =0.0124 $\mu s^{-1}$, and is removed in Fig.~\ref{fig:muSR}. This model is also used for longitudinal field measurements (see Fig. \ref{fig:muSR}.d).

Higher temperature measurements from 2 K to 200 K were performed on TRIUMF's M20 beamline, utilizing an ultra-low background probe (see Fig. \ref{fig:muSR}.a). In this case, the depolarization is modeled as: 

\begin{equation}
    P =  e^{-( \lambda t) ^{\beta}}
    \label{Eqn:M20Depol}
\end{equation}
which corresponds to Eq.~\ref{Eqn:M15Depol} without the silver background (i.e., F=1). The sample fraction of the ZF M15 data was determined globally (that is, determined by the entire ZF M15 data set) to be 0.63. This value was then used for the LF data set. $\beta$, determined as a global parameter of the M15 zero-field temperature-dependent spectra, was determined to be 0.567(4). A similar treatment of the isothermal longitudinal field sweep conducted at 280~mK results in $\beta$=0.516(5). These values are fairly close to $\frac{1}{2}$, which is expected in systems with dilute moments and fast fluctuations \cite{blundell2022muon}. In this case, the fluctuation rate, $\nu$, is inversely proportional to $\lambda$. As such, $\beta$ was held fixed at $\frac{1}{2}$ for the M15 ZF data set. The ZF M20 data was fitted with a temperature-dependent $\beta$, which increased monotonically from 0.509(5) at 2 K to 0.874(6) at 200 K (see Fig.~\ref{fig:muSR}c). This accompanies a monotonically decreasing $\lambda$, which reaches $\sim$ 0.3 $~\mu$s$^{-1}$ at 200~K (see Fig.~\ref{fig:muSR}c). At high temperatures, thermal energy overcomes the influence of the crystal electric fields, and the electronic spin fluctuations are fast. As the temperature decreases, the crystal electric fields alter the density of excitations, and the loss of thermal energy slows the electronic spin fluctuations, leading to a substantial increase in the observed relaxation rate. The polarization decays to zero in our zero-field measurements performed on M20/M15, implying that the relaxation reflects dynamical (fluctuating) fields.  Furthermore, measurements performed with a longitudinal field reveal that relaxation is largely independent of the field up to and including 0.3~T (see Fig.~\ref{fig:muSR}e). In the case where relaxation is due to static moments, the width of a static random field distribution is estimated simply as $\frac{\lambda}{\gamma_{\mu}}$, where $\gamma_{\mu}$ is the gyromagnetic ratio of a muon. A relaxation rate of  $\sim 4\;\mu $s$^{-1}$ would gives an approximate field width of $\sim$3~mT. This further implies a dynamical nature to the relaxation; hence, this field width would be almost fully decoupled by 0.3~T. Interestingly, the relaxation rate actually increases above 0.3~T; this is atypical (dynamical fluctuations eventually decouple, and hence the relaxation rate is expected to decrease with sufficient field) and indicates some combination of a field-induced change in moment size, a slowing down of fluctuations, and / or a change in the density of excitations, as observed in Tb$_2$Sn$_2$O$_7$ \cite{de2006spin}. Lastly, the zero-field $\mu$SR measurements host a plateau in the temperature dependence of the relaxation. The onset of the plateau feature corresponds to roughly the same temperature as the broad anomaly seen in the heat capacity and remains largely temperature-independent below 200~mK.  A plateau in the relaxation rate indicates persistent spin dynamics, which are expected in a QSL,  while the temperature independence implies the fluctuations are quantum in nature. Such plateaus have been observed in numerous other candidate QSL materials (e.g. \cite{FakB2012Kakq, li2016muon, MendelsP2007Qmit}).

\section{Conclusion}

We find that Yb$_2$Be$_2$GeO$_7$, a new SSL material with rare-earth ions on the SSL sites, exhibits no sign of long-range magnetic order or spin freezing in specific heat capacity, magnetometry, PND, nor $\mu$SR. PCCEF calculations reveal an effective S=$\frac{1}{2}$ ground-state, which is reflected in magnetic susceptibility and entropy. Diffuse scattering is observed in the PND pattern, indicating the presence of short-range magnetic correlations. Dynamical spins are observed via $\mu$SR, which persist to the lowest measured temperature of 17~mK. The combination of an absence of magnetic ordering and persistent dynamical spin fluctuations suggests the ground-state of Yb$_2$Be$_2$GeO$_7$ could be a quantum spin liquid.

Further exploration of Yb$_2$Be$_2$GeO$_7$ would be of interest to the field of frustrated magnetism. In particular, inelastic neutron scattering studies are required to elicit (experimentally) the crystal electric fields that are present, to study the zero-field gap of $\sim$0.8 K observed in the specific heat capacity, and to test for possible spinion excitations that are associated with QSLs. Tuning of the $J^{'}$/$J$ ratio via application of pressure may reveal additional phases, as seen in SrCu$_2$(BO$_3$)$_2$. Electron spin resonance can determine the potential role of dilute defects if they are present. 

\section{Acknowledgements}

We thank Dr. Bruce Gaulin and Dr. Jeff Rau for their helpful discussions. Work at McMaster University was supported by the Natural Sciences and Engineering Research Council of Canada. A portion of this research used resources at the High Flux Isotope Reactor, a DOE Office of Science User Facility operated by the Oak Ridge National Laboratory. Work at the University of Birmingham was supported by the UKRI Engineering and Physical Sciences Research Council (EP/V028774/1). The authors also acknowledge support from the University of Birmingham and McMaster University through the BIRMAC Quantum Materials Fund. 

\section{Appendix A: Point Charge Crystal Electric Field Model}

Point charge crystal electric field calculations were performed using the Python package PyCrystalField \cite{scheie2021pycrystalfield}. The structural details (e.g., lattice parameters) used in these calculations were derived via XRD Rietveld refinement (see Table~\ref{Tab:Rietveldpara} of Appendix B). The eight nearest-neighbor oxygen ligands were considered in the calculation.

\begin{figure*}
  \begin{minipage}{.99\linewidth}
    \begin{ruledtabular}
    \begin{tabular}{c|cccccccc}
    E (meV) &$| -\frac{7}{2}\rangle$ & $| -\frac{5}{2}\rangle$ & $| -\frac{3}{2}\rangle$ & $| -\frac{1}{2}\rangle$ & $| \frac{1}{2}\rangle$ & $| \frac{3}{2}\rangle$ & $| \frac{5}{2}\rangle$ & $| \frac{7}{2}\rangle$ \tabularnewline
    \hline 
    0.000 & (0.0231+0j) & -0.0534j & (0.0615+0j) & -0.1325j & (0.1283+0j) & -0.3538j & (0.4205+0j) & -0.8104j \tabularnewline
    0.000 & (0.8104+0j) & 0.4205j & (0.3538+0j) & 0.1283j & (0.1325+0j) & 0.0615j & (0.0534+0j) & 0.0231j \tabularnewline
    31.516 & (-0.42+0j) & 0.48j & (0.0005+0j) & 0.404j & (0.3058+0j) & 0.2886j & (0.4981+0j) & 0.0713j \tabularnewline
    31.516 & (0.0713+0j) & -0.4981j & (0.2886+0j) & -0.3058j & (0.404+0j) & -0.0005j & (0.48+0j) & 0.42j \tabularnewline
    58.002 & (-0.2945+0j) & -0.0309j & (0.6463+0j) & -0.0524j & (0.3615+0j) & 0.2299j & (-0.4962+0j) & -0.2494j \tabularnewline
    58.002 & (-0.2494+0j) & 0.4962j & (0.2299+0j) & -0.3615j & (-0.0524+0j) & -0.6463j & (-0.0309+0j) & 0.2945j \tabularnewline
    117.993 & 0j & 0.2875j & (-0.1249+0j) & -0.7466j & (-0.1358+0j) & 0.5492j & (0.1095+0j) & -0.1108j \tabularnewline
    117.993 & (0.1108+0j) & 0.1095j & (-0.5492+0j) & -0.1358j & (0.7466+0j) & -0.1249j & (-0.2875+0j) & 0j \tabularnewline
    \end{tabular}\end{ruledtabular}
    \captionof{table}{Eigenvectors and Eigenvalues of the CEF Hamiltonian. }
    \label{tab:Yb_Eigenvectors}
  \end{minipage}
  \begin{minipage}[b]{.48\linewidth}
    \raggedright
    \includegraphics[width=0.95\linewidth]{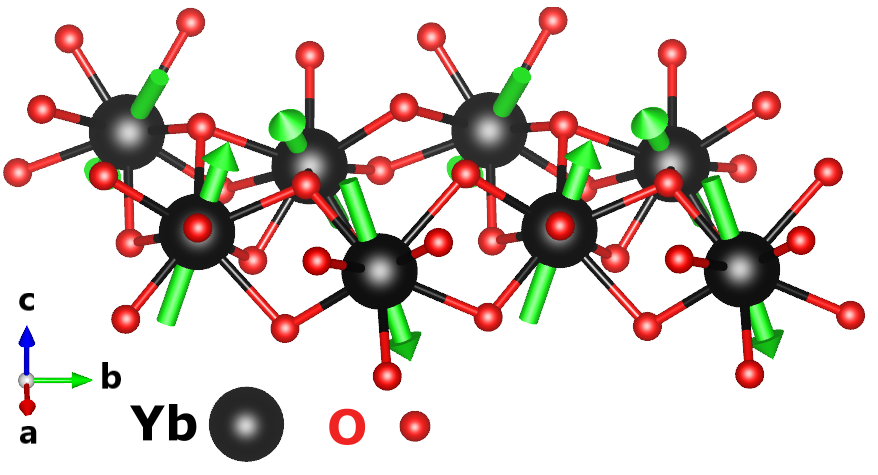}
    \captionof{figure}{Planar Yb$^{3+}$ lattice, with nearest neighbor oxygen ligands, showing the PCCEF local z-axis (green arrows).}
    \label{fig:YbSpinDir}
  \end{minipage}\hfill
  \begin{minipage}[b]{.48\linewidth}
    \centering
    \begin{ruledtabular}
    \begin{tabular}{c|ccc}
    G$_{ij}$ & x& y& z \tabularnewline
    \hline
    x & 0 & -1.3152 &0 \tabularnewline
    y & -0.5983 & 0 & -1.1830 \tabularnewline
    z & -0.1593 & 0 & 6.6594
    \end{tabular}
    \end{ruledtabular}

    \captionof{table}{g-tensor determined via the PCCEF model. The dominant G$_{zz}$ element implies Ising anisotropy along the local z direction.}
    \label{tab:Gtensor}
  \end{minipage}
\end{figure*}

The model predicts Ising anisotropy (see Fig.~\ref{fig:YbSpinDir} and Table~\ref{tab:Gtensor} for details), with a powder magnetization that reaches a moment of 1.61~$\mu_B$/Yb$^{3+}$ in a field of 7 T at 0.4 K. The susceptibility, fit with the two-level model for comparative purposes, is characterized by the parameters $\mu_0$~=~3.6$~\mu_B$, $\mu_1$~=~5.4$~\mu_B$ (see Fig.~\ref{fig:SQUID}), giving high-temperature and low-temperature effective moments of 4.6$~\mu_B$ and 3.6$~\mu_B$ respectively. The PCCEF model also predicts a Kramer's doublet ground-state, mostly occupied by the J=$\pm$7/2 manifold, which is separated from the first excited state by ~32 meV (see Table~\ref{tab:Yb_Eigenvectors}); this is consistent with an effective spin $\frac{1}{2}$ system. Effective spin =$\frac{1}{2}$ systems are commonly realized in materials, across numerous space groups, containing Yb$^{3+}$ ions \cite{PhysRevB.104.024427, PhysRevX.1.021002, li2015gapless} due to the combination of CEFs and spin-orbit coupling \cite{PhysRevB.94.035107}, which conspire to form a J= $\frac{1}{2}$ Kramers's doublet ground-state. 

\section{Appendix B: Rietveld Refinement}

A summary of the Rietveld refinement performed on Yb$_2$Be$_2$GeO$_7$ for both X-ray diffraction (at 300~K) and powder neutron diffraction (at 290~mK). The Rietveld parameters are tabulated in Table. \ref{Tab:Rietveldpara}. Graphical representation of the data set is presented in the main body, under Fig.~\ref{fig:elastic} of the results section. The refinement was performed using FullProf. The space group P$\overline{\mbox{4}}$2$_1$m (113) was used for both refinements.

\begin{table*}
\begin{ruledtabular}
    \begin{tabular}{c|c|cccc|c|ccc}
    PND & Atom & x& y& z &Occ. &Phase &a, \AA&b, \AA& c, \AA \tabularnewline
    \hline
     &Yb & 0.1583(3) & 0.3417(3) & 0.5035(7) & 0.5& {Yb$_2$Be$_2$GeO$_7$} &7.3005(9) &7.3005(9)& 4.7351(6) \tabularnewline
     &Be & 0.1354(4)& 0.6354(4) & 0.0479(8) &0.50 & &  & &  \tabularnewline
     &Ge & 0.0 & 0.0 & 0.0 & 0.250  &  &R$_p$ &R$_{wp}$ & $\lambda$, \AA \tabularnewline
     &O$_1$ & 0.0832(5) & 0.83315(6) & 0.218(1) &1 & & 8.60 & 12.9& 1.486 \tabularnewline
     &O$_2$ & 0.1409(5) & 0.6409(5) & 0.716(1) &0.50 &  &  & & \tabularnewline
     &O$_3$ & 0 & 0.5 & 0.187(2) & 0.25 &  & &  & \tabularnewline
    \hline
     XRD&Yb & 0.15831(8) & 0.34169(8) & 0.507(7) & 0.5&Phase &a, \AA&b, \AA& c, \AA \tabularnewline
     &Be & 0.134(2)& 0.634(2) & 0.021(7) &0.50 & Yb$_2$Be$_2$GeO$_7$ &7.33557(2) &7.33557(2)& 4.75520(2) \tabularnewline
     &Ge & 0.0 & 0.0 & 0.0 & 0.250 &GeO$_2$ & 4.34980(1) & 4.34980(1)& 2.79704(9) \tabularnewline
     &O$_1$ & 0.084(1) & 0.833(1) & 0.200(1) &1 &Yb$_2$O$_3$&10.2600(1) & 10.2600(1)& 10.2600(1)\tabularnewline 
     &O$_2$ & 0.144(1) & 0.644(1) & 0.731(2) &0.50 &  &R$_p$ &R$_{wp}$ & $\lambda$, \AA \tabularnewline
     &O$_3$ & 0 & 0.5 & 0.178(3) & 0.25 &  & 13.8 & 21.7& 1.541 \tabularnewline
    \end{tabular}
\end{ruledtabular}
\caption{Summary of the Rietveld refinement parameters of the X-ray diffraction (Cu K$_{\alpha1}$ $\lambda=1.541$~\AA, T = 300 K) and powder neutron diffraction (Ge (113) $\lambda$=1.486~\AA, T= 290 mK). The parameters \{x, y, z\} and \{a, b, c\} are the fractional coordinates and lattice parameters, respectively. Error-less values were not refined.}
\label{Tab:Rietveldpara}
\end{table*}

\newpage

\end{document}